\documentclass[aps,pra,twocolumn,longbibliography,showpacs,showkeys,amsmath,superscriptaddress,floatfix]{revtex4-1}
\usepackage{graphicx}
\usepackage{epsfig}
\usepackage{dcolumn}
\usepackage{bm}
\usepackage{amsmath}
\usepackage{amsbsy}
\usepackage{amssymb}
\usepackage[usenames]{color}
\usepackage{multirow}
\usepackage{tabu}
\usepackage{soul}

\begin{document}


\title{Bessel beams revisited: a generalized scheme to derive optical vortices}

\author{G. F. Quinteiro}

\email{gquinteiro@df.uba.ar}

\affiliation{ Departamento de F\'isica and IFIBA, FCEN, Universidad de Buenos Aires,
Ciudad Universitaria, Pabell\'on I, 1428 Ciudad de Buenos Aires, Argentina}

\author{C. T. Schmiegelow}

\affiliation{ Departamento de F\'isica and IFIBA, FCEN, Universidad de Buenos Aires,
Ciudad Universitaria, Pabell\'on I, 1428 Ciudad de Buenos Aires, Argentina}


\author{D. E. Reiter}

\affiliation{Universit\"at M\"unster, Wilhelm-Klemm-Str. 10, 48149 M\"unster,
Germany}

\author{T. Kuhn}

\affiliation{Universit\"at M\"unster, Wilhelm-Klemm-Str. 10, 48149 M\"unster,
Germany}

\date{\today}

\begin{abstract}
The electromagnetic field of optical vortices is in most cases derived from vector and scalar potentials using either a procedure based on the Lorenz or the Coulomb gauge. The former procedure has been typically used to derive paraxial solutions with Laguerre-Gauss radial profiles, while the latter procedure has been used to derive full solutions of the wave equation with Bessel radial profiles. We investigate the differences in the derivation procedures applying each one to both Bessel and Laguerre-Gauss profiles. We show that the electromagnetic fields thus derived differ in the relative strength of electric and magnetic contributions. The new solution that arises from the Lorenz procedure in the case of Bessel beams restores a field symmetry that previous work failed to resolve. Our procedure is further generalized and we find a spectrum of fields beyond the Lorenz and Coulomb gauge types. Finally, we describe a possible experiment to test our findings.
\end{abstract}

\maketitle


\section{Introduction}

Electromagnetic waves in free space are transverse waves, as determined by
Maxwell's divergence equations $\nabla \cdot \mathbf{E} = 0$ and $\nabla
\cdot \mathbf{B} = 0$. For the simplest case of plane waves, this leads to
the fact that both the electric and the magnetic field have no field component
along the propagation direction. The same holds for spherical waves in the
far-field zone emitted from localized charge distributions such as, e.g.,
electric or magnetic dipole radiation. For light beams with a finite beam
diameter or for beams with a more complicated phase distribution in general
the electric and the magnetic field may have components along the propagation
direction. Nevertheless, these components are often smaller than the
components perpendicular to the propagation direction. Then, to obtain the
full electromagnetic fields of these beams it may be a good strategy to start
with an ansatz for the components transverse to the propagation direction and
to construct from these the full fields in such a way that they satisfy
Maxwell's equations.

In electrodynamics it is often convenient to derive the electromagnetic
fields from the vector potential $\mathbf{A}$ and the scalar potential $\Phi$
\cite{jackson1999classical}. As is well known, these potentials are not
uniquely determined, instead there is a gauge freedom. Probably the most
common gauges are the Lorenz and the Coulomb gauge. For plane waves,
both gauges are simultaneously satisfied for a vector potential that, like
the electromagnetic fields, has only components perpendicular to the
propagation direction and a vanishing scalar potential. Again, for beams with
finite beam diameter or a more complicated phase distribution this does not
hold anymore, but it can be taken as a starting point to complement the
transverse components of $\mathbf{A}$ by a longitudinal one and/or a scalar
potential in such a way that the Coulomb or Lorenz gauge condition is
satisfied.

A class of light beams with finite lateral extension which has attracted
large interest in the past years are optical vortices (OV) or twisted light
beams, i.e., light fields exhibiting phase singularities. Such beams are interesting because of their
possibility to carry orbital angular momentum (OAM) which could be exploited
for applications in the field of quantum information technology
\cite{molina2001management,bozinovic2013ter, ren2016chip, mirhosseini2013efficient, wang2012terabit, krenn2014communication}.
Among those beams, Laguerre-Gaussian (LG) \cite{loudon2003theory,
romero2002qua, allen1992orb} and Bessel \cite{volke2002orb, jauregui2004rot, matula2013atomic} beams are the most widely considered types. Interestingly, for these two types of beams
different strategies have been used in the literature to complement the
transverse components of the vector potential. In the case of LG modes, the
point of view of the Lorenz gauge has been applied \cite{loudon2003theory,
romero2002qua, allen1992orb}: A transverse vector potential with a LG radial
mode is postulated. Then, the Lorenz gauge is imposed to derive a scalar
potential which complements the two components of the vector potential such
that the Lorenz condition is fulfilled, and finally electric and magnetic
fields are derived from these potentials. The electromagnetic fields thus
found satisfy the paraxial wave equation.

The derivation of Bessel beams usually takes a different path
\cite{volke2002orb, jauregui2004rot, matula2013atomic}. Besides the obvious
choice of Bessel type radial modes, the common point is the assumption of a
null scalar potential and the use of the Coulomb gauge to determine the longitudinal component $A_z$ of the vector potential. The electromagnetic fields derived from these potentials satisfy the full wave equation.

The two procedures sketched above differ by the choice of the gauge and the ansatz for the spatial mode. If both procedures were carried out on the same radial function --LG or Bessel--, one might expect that the results are
related by a gauge transformation alone. If this were true, the electromagnetic
fields derived from each procedure would be one and the same. To answer this
and other related questions, in this paper we follow both procedures and
analyze their general properties. We will show that, in contrast to what
might be expected, there is no possible gauge transformation connecting both
sets of potentials, and that they therefore lead to different electromagnetic
fields. We derive a generalized procedure which interpolates smoothly between
these two cases. The general theory is then applied to vortex beams. Here we
discuss in particular the class of Bessel beams because they are exact solutions of
Maxwell's equations (or the full wave equation) and we therefore do not have
to worry about possible different orders in the paraxial approximation
\cite{lax1975maxwell}, which might complicate the comparison. The
same strategy is then applied to LG beams taking into account that they are
solutions of the wave equation in the paraxial approximation. In particular
we will show that the two approaches based on either a vanishing scalar
potential or a vanishing longitudinal component of the vector potential give
rise to two distinct types of electromagnetic fields. These differences, however, are at least of second order in the paraxial parameter and therefore require to go beyond the regime of paraxial beams. Finally we discuss how these differences could be measured in the case of tightly focused Bessel beams. 

The article is organized as follows. In Sec.~\ref{general} the procedure to
construct either a longitudinal component of the vector potential or a scalar
potential is introduced. This procedure is then generalized in such a way to
continuously interpolate between these limiting cases, and the resulting
electric and magnetic fields are discussed. Section~\ref{Vortices} is devoted
to the application of the general formalism to the case of vortex beams. We discuss in particular the case of Bessel beams, which are exact solutions of Maxwell's equations, and then we show that the same features are found for LG beams when going beyond the paraxial regime.
Finally, in Sec.~\ref{experiment} we discuss a possible experiment to test
the new solutions found in the previous sections. We end with some
conclusions, that include a discussion on the light-matter interaction of the
newly-found fields.

\section{General formalism}\label{general}

Expressing the electromagnetic fields in terms of a scalar and a vector
potential and inserting these potentials into Maxwell's equations in general
leads to coupled equations of motion for the potentials. To decouple them,
the gauge freedom can be employed. A common gauge is the Lorenz gauge
\begin{equation}\label{Eq:Lorenz}
\nabla \cdot \mathbf{A}(\mathbf{r},t) + \frac{1}{c^2} \partial_t \Phi(\mathbf{r},t) = 0 .
\end{equation}
Within this gauge, far from any sources all the components of $\mathbf{A}$ as
well as $\Phi$ satisfy the homogeneous wave equation.

Another common gauge is the Coulomb gauge
\begin{equation}\label{Eq:Coulomb}
\nabla \cdot \mathbf{A}(\mathbf{r},t) = 0 ,
\end{equation}
which leads to Poisson's equation for $\Phi$ and a wave equation for
$\mathbf{A}$. Far from the sources the scalar potential vanishes, i.e., $\Phi
= 0$. In this regime the Coulomb gauge is also called radiation gauge, and it
is then a special case of the Lorenz gauge.

For plane waves with wave vector $\mathbf{q}$, the Coulomb gauge requires
$\mathbf{q} \cdot \mathbf {A} = 0$ stating that the vector potential has no
component in the propagation direction. In addition, the radiation condition is imposed and $\Phi=0$. For beams with finite lateral
extensions in general this does not hold anymore. Nevertheless, it is often a
good starting point, e.g., for a monochromatic beam with frequency $\omega$
traveling in $z$-direction, to make the ansatz
\begin{equation}\label{Eq:Aperp}
\mathbf{A}(\mathbf{r},t)=\mathbf{A}_\perp(\mathbf{r},t)
=\tilde{\mathbf{A}}_\perp(\mathbf{r})e^{i(q_z z-\omega t)}
\end{equation}
with $\tilde{\mathbf{A}}_\perp(\mathbf{r})$ lying in the $xy$-plane and
chosen in such a way that the two components of $\mathbf{A}$ satisfy the
homogeneous wave equation. However, assuming $A_z=\Phi=0$ as above,
Eq.~\eqref{Eq:Aperp} in general neither satisfies Coulomb nor Lorenz gauge,
and the fields derived from this potential do not satisfy Maxwell's
equations. This drawback can be corrected in different ways. In the following
we will start by discussing two limiting cases of such an extension of
Eq.~\eqref{Eq:Aperp}, which will then be generalized to a whole class of
potentials.

\subsection{{\it 2CA} and {\it 3CA} potentials}\label{2CA-3CA}

Requesting that the potentials satisfy the Lorenz gauge condition for a completely transverse vector potential with components $A_x$ and $A_y$, a scalar potential can be derived from Eq.~\eqref{Eq:Lorenz}. The potentials then read
\begin{subequations}
\label{Eq:2CA}
\begin{eqnarray}
\label{Eq:2CA_Aperp}
\mathbf{A}_\perp^{(0)}(\mathbf r,t) &=& \mathbf{A}_\perp(\mathbf r,t) , \\
\label{Eq:2CA_Az}
A_z^{(0)}(\mathbf r,t) &=& 0 , \\
\label{Eq:2CA_Phi}
\Phi^{(0)}  (\mathbf r,t) &=& - i \frac{c^2}{\omega} \nabla_\perp
\cdot \mathbf{A}_\perp (\mathbf r, t).
\end{eqnarray}
\end{subequations}
We will refer to this choice of potentials with two non-vanishing components
of the vector potential as {\it 2CA potentials} and for reasons that will
become clear below we denote them by a superscript $(0)$.

Since $\mathbf{A}_\perp$ satisfies the homogeneous wave equation, it
follows from Eq.~\eqref{Eq:2CA_Phi} and from the monochromaticity of the scalar potential that also $\Phi^{(0)}$ satisfies the
homogeneous wave equation which, together with the fact that the potentials
satisfy the Lorenz gauge condition, guarantees that the resulting
electromagnetic fields are solutions of Maxwell's equations.

Alternatively, requesting that the potentials satisfy Coulomb and radiation
gauge, the third component $A_z$ of the vector potential can be obtained from
the Coulomb gauge condition Eq.\ (\ref{Eq:Coulomb}). The potentials then read
\begin{subequations}
\label{Eq:3CA}
\begin{eqnarray}
\label{Eq:3CA_Aperp}
\mathbf{A}_\perp^{(1)}(\mathbf r,t) &=& \mathbf{A}_\perp(\mathbf r,t) , \\
\label{Eq:3CA_Az}
\partial_z A_z^{(1)} (\mathbf{r},t) &=&  -  \nabla_\perp \cdot
\mathbf{A}_\perp(\mathbf{r},t)
 , \\
\label{Eq:3CA_Phi}
\Phi^{(1)}  (\mathbf r,t) &=& 0.
\end{eqnarray}
\end{subequations}
We will refer to this choice of potentials with three components of the
vector potential and vanishing scalar potential as {\it 3CA potentials} and
denote them by a superscript $(1)$. 

Again, the fact that $\mathbf A_\perp$ satisfies the homogeneous wave
equation implies through Eq.~\eqref{Eq:3CA_Az} that also $A_z^{(1)}$ satisfies the homogeneous wave equation. Together with the vanishing
scalar potential and the fact that the vector potential satisfies the Coulomb
gauge condition, this guarantees that the resulting electromagnetic fields
are solutions of Maxwell's equations.

\subsection{Are {\it 2CA} and {\it 3CA} potentials connected by
a gauge transformation?} \label{Sec:gauge}

The derivations above were based on different gauges. Therefore, they might
lead to the impression that both sets of potentials are in fact different
gauge representations of the same electromagnetic fields. To check whether
this is true we are looking for a gauge function $\chi(\mathbf r,t)$ that
transforms the {\it 3CA} potentials into the {\it 2CA} potentials. A general
gauge transformation \cite{jackson1999classical} connecting these potentials
should read
\begin{subequations}
\begin{eqnarray}
\label{Eq:A'}
  \mathbf A^{(0)}(\mathbf r,t)
&=&
  \mathbf A^{(1)}(\mathbf r,t) + \nabla \chi(\mathbf r,t)
\,,
\\
\label{Eq:Phi'}
  \Phi^{(0)}(\mathbf r,t)
&=&
  \Phi^{(1)}(\mathbf r,t) - \partial_t \chi(\mathbf r,t)
\,.
\end{eqnarray}
\end{subequations}
On the one hand we have $A^{(0)}_z=0$, therefore $\chi$ must satisfy
\begin{eqnarray}
\label{Eq:chi_z}
    \partial_z \chi(\mathbf r,t)
&=&
    - A^{(1)}_z(\mathbf r,t)
\,.
\end{eqnarray}
In addition, our assumption of the equality of the transverse components,
i.e., $\mathbf{A}_\perp^{(0)}=\mathbf{A}_\perp^{(1)}$, tells us that
$\partial_x \chi = \partial_y \chi = 0$, which implies that
\begin{eqnarray}
\chi = \eta(z,t)
\end{eqnarray}
with a function $\eta$ which depends only on $z$ and $t$. This contradicts
Eq.~\eqref{Eq:chi_z}, since from Eq.~\eqref{Eq:3CA_Az} we know that
$A^{(1)}_z$ has in-plane dependence.

On the other hand we have $\Phi^{(1)}=0$, such that $\chi$ must satisfy
\begin{eqnarray}
\label{Eq:chi_t}
    \partial_t \chi(\mathbf r,t)
&=&
    -\Phi^{(0)}(\mathbf r,t).
\end{eqnarray}
This again implies that $\chi$ has in-plane dependence, which contradicts
the assumption of the equality of the transverse components of $\mathbf A$.

Therefore, we conclude that there is no gauge transformation connecting the
two set of potentials and, as a consequence, they lead to different sets of
electromagnetic fields.

We want to remark that the crucial point is the assumption that the
transverse components of the vector potentials agree in both gauges. Without
this assumption, a gauge function satisfying Eq.~\eqref{Eq:chi_z} can be used
to remove the $z$-component of $\mathbf{A}^{(1)}$ and create instead a scalar
potential. However, the transformed vector potential will then have modified
$x$- and $y$-components. The same holds if a gauge function satisfying
Eq.~\eqref{Eq:chi_t} is used to remove the scalar potential $\Phi^{(0)}$ by
creating a $z$-component of $\mathbf{A}$.

\subsection{Generalized potentials}\label{generalized}

Since {\it 2CA} and {\it 3CA} potentials are not connected by a gauge
transformation, the question arises whether the two procedures discussed
above might be seen as limiting cases of a more general approach to construct
complete potentials from the given in-plane components of
Eq.~\eqref{Eq:Aperp}. For that purpose we introduce an arbitrary real number
$\gamma$ and define the potentials according to
\begin{subequations}
\label{Eq:gamma}
\begin{eqnarray}
\label{Eq:gamma_Aperp}
\mathbf{A}_\perp^{(\gamma)}(\mathbf r,t) &=& \mathbf{A}_\perp(\mathbf r,t) , \\
\label{Eq:gamma_Az}
\partial_z A_z^{(\gamma)} (\mathbf{r},t) &=& - \gamma \nabla_\perp \cdot
\mathbf{A}_\perp (\mathbf{r},t) , \\
\label{Eq:gamma_Phi}
\Phi^{(\gamma)}(\mathbf{r},t) &=& - i\left( 1-\gamma \right)\frac{c^2}{\omega} \nabla_\perp \cdot
\mathbf{A}_\perp(\mathbf{r},t).
\end{eqnarray}
\end{subequations}
For $\gamma=0$ we recover the {\it 2CA} potentials with
$A_z^{(0)}=0$ and for $\gamma=1$ we obtain the {\it 3CA} potentials with
$\Phi^{(1)}=0$. Again, due to the fact that
$\mathbf{A}_\perp^{(\gamma)} = \mathbf{A}_\perp$
satisfies the wave equation, also $A_z^{(\gamma)}$ and $\Phi^{(\gamma)}$
satisfy the wave equation. By construction the potentials
$\mathbf{A}^{(\gamma)}$ and $\Phi^{(\gamma)}$ fulfill the Lorenz gauge
condition \eqref{Eq:Lorenz}. This ensures that the fields calculated from
these potentials are indeed solutions of Maxwell's equations.

\subsection{Electric and magnetic fields}\label{fields}

Given the scalar and the vector potential, the electric and magnetic fields
are determined from
\begin{subequations}
\label{Eq:EB_pot}
\begin{eqnarray}
\label{Eq:E_po}
  \mathbf E(\mathbf r, t) &=& -\partial_t \mathbf A(\mathbf r, t) - \nabla
\Phi(\mathbf r, t)\,, \\
\label{Eq:B_pot}
\mathbf B(\mathbf r, t) &=& \nabla \times \mathbf
A(\mathbf r, t)\,.
\end{eqnarray}
\end{subequations}
Starting from Eq.~\eqref{Eq:gamma} we note that while $\Phi^{(\gamma)}$ is explicitly given in terms of the transverse components of the vector potential, $A_z^{(\gamma)}$ is only given up to an integration. This is because in general $\tilde{\mathbf{A}}_\perp(\mathbf{r})$ in Eq.~\eqref{Eq:Aperp} may depend on
$z$; therefore Eq.~\eqref{Eq:gamma_Az} cannot be explicitly integrated without
specifying $\tilde{\mathbf{A}}_\perp(\mathbf{r})$. The characteristic length scale for the variation of $\tilde{\mathbf{A}}_\perp(\mathbf{r})$ along $z$ is given by the diffraction length $l$ of the beam \cite{lax1975maxwell}. Non-diffracting beams, such as Bessel beams, have an infinite diffraction length and the derivative in Eq.~\eqref{Eq:gamma_Az} is simply given by $\partial_z A_z^{(\gamma)} = i q_z A_z^{(\gamma)}$. Gaussian-like beams, such as Laguerre-Gaussian or Hermite-Gaussian beams are characterized by a finite value of the diffraction length; however, since the diffraction length is typically much larger than the wavelength of the light, it is still a good approximation to set $\partial_z A_z^{(\gamma)} \approx i q_z A_z^{(\gamma)}$. Below, when discussing the application of the formalism to the case of LG beams, we will quantitatively estimate this approximation in terms of the paraxial parameter. 
To get explicit formulas, here we will assume $\partial_z A_z = i q_z A_z$. 

Introducing $\overline{\gamma}=1-\gamma$, this leads to
\begin{subequations}\label{Eq:EB}
\begin{eqnarray}
\label{Eq:E}
  \mathbf E^{(\gamma)}(\mathbf r, t)
&=&
  i \left[
    \omega A_x
    + \overline{\gamma} \frac{c^2}{\omega}  \, \partial_x (\nabla_\perp \cdot \mathbf A_\perp)
  \right] \mathbf{e}_x
\nonumber \\
&&+
  i \left[
    \omega A_y
    + \overline{\gamma} \frac{c^2}{\omega}  \,\partial_y (\nabla_\perp \cdot \mathbf A_\perp)
  \right] \mathbf{e}_y
\nonumber \\
&&-
  \left[
    \gamma \frac{\omega}{q_z}
    + \overline{\gamma} \frac{c^2 q_z}{\omega}
  \right]
  \,(\nabla_\perp \cdot \mathbf A_\perp) \mathbf{e}_z \,, \mbox{\qquad}
\\
\label{Eq:B}
  \mathbf B^{(\gamma)}(\mathbf r, t)
&=&
  - \left[
    \partial_z A_y
    -i \gamma \frac{1}{q_z} \partial_y (\nabla_\perp \cdot \mathbf A_\perp)
  \right] \mathbf{e}_x
\nonumber \\
&&+
  \left[
    \partial_z A_x
    - i \gamma \frac{1}{q_z} \partial_x (\nabla_\perp \cdot \mathbf A_\perp)
  \right] \mathbf{e}_y
\nonumber \\
&&+
  (\partial_x A_y - \partial_y A_x)  \mathbf{e}_z
\,,
\end{eqnarray}
\end{subequations}
where the arguments of the vector potential have been omitted to make the
formulas clearer.

Indeed we find that the electromagnetic fields depend on the value of
$\gamma$, i.e., the different procedures to complement the transverse
components of the vector potential give rise to different electromagnetic
fields. These differences are all related to the term $\nabla_\perp \cdot
\mathbf{A}_\perp$. Interestingly, in the limiting cases of {\it 2CA} (i.e.,
$\gamma=0$) and {\it 3CA} (i.e., $\gamma=1$) these correction terms appear
either in the transverse components of the electric field or of the magnetic
field. We will come back to the consequences of these corrections below. The
$z$-component of the magnetic field is unaffected by these terms while the
$z$-component of the electric field is completely caused by $A_z$ and $\Phi$.

\section{Application to optical vortices}\label{Vortices}

In this section we will apply the formalism to the two most important types of OV, namely Bessel beams and LG beams. Beams with a finite diameter are typically classified in terms of a so-called paraxial parameter $f$, which is given by the ratio of the wavelength of the light to a characteristic width of the beam. For Bessel beams the beam diameter can be expressed as an inverse transverse wave vector $q_r$, and the paraxial parameter is then given by  $f=q_r/q_z$ with $q_z$ denoting the wave vector component in propagation direction. LG beams are characterized by a beam waist $w_0$ and the paraxial parameter is defined as $f=(q_z w_0)^{-1}$. 

Bessel beams provide an exact solution of Maxwell's equations or,
equivalently, the full wave equation without performing a paraxial
approximation. They are obtained from a factorization in cylindrical
coordinates and they constitute an example for the class of non-diffracting
beams, i.e., beams with a transverse profile that is independent of the
coordinate along the propagation direction. While the paraxial parameter $f$ is a useful quantity to estimate the importance of certain contributions, as will be done below, it should be noted that all the results are valid at any order of $f$.

LG beams are solutions of the paraxial wave equation, which is obtained from the full wave equation by neglecting terms of the order $f^2$. Thus strictly speaking only contributions up to first order in $f$ should be kept and all higher order corrections should be neglected to remain consistent with the paraxial wave equation. In particular, since the diffraction length of an LG beam is given by $l=q_z w_0^2$ \cite{lax1975maxwell} we have $(q_z l)^{-1} =f^2$, 
which shows that the approximation $\partial_z A_z = i q_z A_z$ used in the previous section is indeed correct up to corrections of the order $f^2$. However, also LG beams can be considered beyond the paraxial limit either by using a systematic expansion of the wave equations in powers of $f$ \cite{lax1975maxwell, davis1979theory, couture1981gaussian, agrawal1983free} or by explicitly including the focusing of LG beams by a lens with high numerical aperture \cite{monteiro2009ang, bliokh2011spin, iketaki2007inv}.

In the following we will first study in detail the case of Bessel beams and then briefly analyze the case of LG beams.

\subsection{Bessel beams}\label{Bessel}

The transverse components of the vector potential for a Bessel beam with well-defined orbital angular momentum
(OAM) and spin angular momentum (SAM) is given by
\cite{quinteiro2015formulation}
\begin{eqnarray}\label{Eq:OV}
\mathbf A_\perp(\mathbf r,t)
=  A_0  J_\ell(q_r r) e^{i \ell \varphi }
e^{i (q_z z -\omega t)} {\boldsymbol \epsilon}_\sigma ,
\end{eqnarray}
where $ J_\ell(q_r r)$ is a Bessel functions of the first kind of
order $\ell$, $q_r^{-1}$ characterizes the beam waist, ${\boldsymbol \epsilon}_\sigma = (\mathbf{e}_x + i \sigma
\mathbf{e}_y)/\sqrt{2}$ with $\sigma=\pm 1$ is the polarization vector for
circular polarization and the integer $\ell$ is the topological index. Such a field carries OAM and
SAM per photon of $\hbar \ell$ and $\hbar \sigma$, respectively. With
$\omega^2 = c^2 \left(q_z^2 + q_r^2 \right)$ the two components of this
vector potential satisfy the wave equation, however, the potential does
neither satisfy the Lorenz nor the Coulomb gauge condition. Therefore, the
goal of the following subsections will be to complement these transverse
components by a $z$-component $A_z$ and/or a scalar potential $\Phi$.

\subsubsection{Potentials}

We now apply the general procedure Eqs.~\eqref{Eq:gamma} to the special case of Bessel beams. All the correction terms are related to the term $\nabla_\perp \cdot \mathbf{A}_\perp$. Using
Eq.~\eqref{Eq:OV}, this term reads
\begin{eqnarray}
\hspace{-6mm}  
  \nabla_\perp \cdot \mathbf A_\perp(\mathbf r, t)
&=&
  - \sigma \frac{A_0}{\sqrt{2}}  q_r
  J_{\ell+\sigma}(q_r r)  e^{i (\ell + \sigma) \varphi}
  e^{i (q_z z -\omega t)}.
\end{eqnarray}
The potentials then read
\begin{subequations}
\begin{eqnarray}
\label{Eq:gamma_vectpot}
    \mathbf {\tilde{A}}^{(\gamma)}(\mathbf r)
&=&
    A_0 J_\ell(q_r r) e^{i \ell \varphi }
    {\boldsymbol{\epsilon}}_\sigma
\nonumber \\
&&
    -i \gamma \sigma \frac{q_r}{q_z} \frac{A_0}{\sqrt{2}} J_{\ell+\sigma}(q_r r)
    e^{i (\ell+\sigma) \varphi }
    {\mathbf{e}}_z,
\\
\label{Eq:gamma_scalarpot}
    {\tilde{\Phi}}^{(\gamma)}(\mathbf r)
&=&
    i \overline{\gamma} \frac{c^2}{\omega} \sigma \frac{A_0}{\sqrt{2}} q_r
    J_{\ell+\sigma}(q_r r) e^{i (\ell+\sigma) \varphi }
\end{eqnarray}
\end{subequations}
For $\gamma=1$ (i.e., $\overline{\gamma}=0$) we obtain the {\it 3CA}
potentials for Bessel beams, which is the form typically used for this kind
of beams \cite{jauregui2004rot, quinteiro2015formulation}. On the other hand,
for $\gamma=0$ (i.e., $\overline{\gamma}=1$) we get the {\it 2CA} potentials,
which is the form that has been used in the literature for LG beams
\cite{loudon2003theory, romero2002qua, allen1992orb}. 

We notice that both $A_z$ and $\Phi$ have the paraxial parameter $f=(q_r/q_z)$ or powers of it as a prefactor, which demonstrates that they are indeed smaller than the transverse components $\mathbf{A}_\perp$ for collimated beams.
In the following we will analyze in detail the differences in the
electromagnetic fields of Bessel beams as obtained from the {\it 2CA} or {\it
3CA} potentials.

\subsubsection{Electric and magnetic fields}

We gain further insight by writing the fields in the basis of circular
polarization given by the basis vectors $\mathbf{e}_\sigma=(\mathbf{e}_x + i \sigma \mathbf{e}_y)/\sqrt{2}$ and
$\mathbf{e}_z$. This is done by using the relations $\partial_x =
\cos(\varphi) \partial_r - (1/r) \sin(\varphi)
\partial_\varphi$ and $\partial_y = \sin(\varphi) \partial_r + (1/r)
\cos(\varphi) \partial_\varphi$ and projecting the in-plane components of the
fields on the unit vectors $\mathbf{e}_\pm = \mathbf{e}_{\sigma=\pm 1}$. To simplify the
notation, we write the fields in the form $\mathbf E(\mathbf r,t) =
\tilde{\mathbf E}(\mathbf r) e^{i (q_z z -\omega t)}$ and
$\mathbf B(\mathbf r,t) = \tilde{\mathbf B}(\mathbf r) e^{i (q_z z -\omega
t)}$.
Then we obtain in the {\it 2CA} limit the fields
\begin{subequations}
\begin{eqnarray}
\label{Eq:E_2CA_b}
  \tilde{\mathbf E}^{(0)}(\mathbf r)
&=&
  i E_ 0 J_\ell (q_r r) e^{i \ell \varphi}
  {\mathbf{e}}_\sigma
  + i \sigma \frac{E_0}{\sqrt{2}} \left(\frac {c q_r}{\omega}\right)^2
\nonumber \\
&& \times
\left[
  J_{\ell + \sigma + 1}(q_r r) e^{i (\ell + \sigma + 1) \varphi}
  {\mathbf{e}}_-
\right.
\nonumber \\
&&
\left.
  - J_{\ell + \sigma - 1} (q_r r) e^{i (\ell + \sigma - 1) \varphi}
   {\mathbf{e}}_+
   \right]
\nonumber \\
&&+
   \sigma c^2 \frac {q_z q_r} {\omega^2} \frac{E_0}{\sqrt{2}}
   J_{\ell + \sigma}(q_r r)
   e^{(\ell + \sigma)\varphi} {\mathbf{e}}_z , \\
\label{Eq:B_2CA_b}
  \tilde{\mathbf B}^{(0)}(\mathbf r)
&=&
  \sigma B_0 J_\ell(q_r r) e^{i \ell \varphi}
         {\mathbf{e}}_\sigma
\nonumber \\
&&-
  i \frac{B_0}{\sqrt{2}} \frac{q_r}{q_z} J_{\ell + \sigma} (q_r r)
  e^{i (\ell + \sigma) \varphi} {\mathbf{e}}_z
\,.
\end{eqnarray}
\end{subequations}
In the {\it 3CA} case the fields read
\begin{subequations}
\begin{eqnarray}
\label{Eq:E_3CA_b}
  \tilde{\mathbf E}^{(1)}(\mathbf r)
&=&
  i E_ 0 J_\ell (q_r r) e^{i \ell \varphi}
  {\mathbf{e}}_\sigma
\nonumber \\
&&+
   \sigma \frac {q_r}{q_z} \frac{E_0}{\sqrt{2}} J_{\ell + \sigma}(q_r r)
   e^{(\ell + \sigma)\varphi} {\mathbf{e}}_z , \\
\label{Eq:B_3CA_b}
  \tilde{\mathbf B}^{(1)}(\mathbf r)
&=&
  \sigma B_0 J_\ell(q_r r) e^{i \ell \varphi}
         {\mathbf{e}}_\sigma
  + \sigma \frac{B_0}{\sqrt{2}} \left(\frac {c q_r}{\omega}\right)^2
\nonumber \\
&& \times
\left[
  J_{\ell + \sigma + 1}(q_r r) e^{i (\ell + \sigma + 1) \varphi}
  {\mathbf{e}}_-
\right.
\nonumber \\
&&
\left.
  + J_{\ell + \sigma - 1} (q_r r) e^{i (\ell + \sigma - 1) \varphi}
   {\mathbf{e}}_+
   \right]
\nonumber \\
&&-
  i \frac{B_0}{\sqrt{2}} \frac{q_r}{q_z} J_{\ell + \sigma} (q_r r)
  e^{i (\ell + \sigma) \varphi} {\mathbf{e}}_z
\,,
\end{eqnarray}
\end{subequations}
where we have used $B_0 = q_z A_0$ and $E_0 = \omega A_0$. In the following
we will discuss the differences between the fields resulting from the {\it
2CA} and {\it 3CA} potentials. Note that these formulas are exact due to the non-diffracting nature of Bessel beams.

Let us first concentrate on the components along the propagation direction. As already mentioned, the magnetic field component $B_z^{(\gamma)}$ is independent of the parameter $\gamma$ and therefore the same in all cases. The spatial profile of the electric field component $E_z^{(\gamma)}$ is also the same for all values of $\gamma$, however the prefactor slightly depends on $\gamma$. If we expand the prefactor in powers of the paraxial parameter $(q_r/q_z)$ we find that the lowest order, i.e., the linear term, is independent of $\gamma$ while the next order, which is  $\sim (q_r/q_z)^3$, depends on the value of $\gamma$. This is in agreement with the findings of Lax {\it{et al.}}~\cite{lax1975maxwell}, who showed that when performing the expansion in the paraxial parameter the field components in propagation direction have only contributions of odd orders while the transverse components have only even orders. Furthermore, the $n$-th order of the longitudinal components is completely determined by the $(n-1)$-th order of the transverse components, which is the reason why  the lowest order is independent of $\gamma$.   

Let us now come to the transverse components. When looking at Eqs.~\eqref{Eq:E_2CA_b} and \eqref{Eq:E_3CA_b}, we observe that in both cases the transverse components of one of the two field types -- the electric or the magnetic one -- keep the circular polarization of the transverse components of the vector
potential, while the other one gets an additional contribution with the
opposite circular polarization. Instead of using the classification in terms
of {\it 2CA} and {\it 3CA} potentials, which is based on the unmeasurable
potentials, we can therefore introduce a classification in terms of the
fields and thus a classification which is directly related to measurable
quantities. In analogy with transverse electric (TE) and transverse magnetic
(TM) fields in the case of guided waves, we can classify the fields in the
{\it 3CA} case as {\it circular electric} (CE) and those in the {\it 2CA}
case as {\it circular magnetic} (CM) because in the former case the
transverse components of the electric field have a well-defined circular
polarization while in the latter case this holds for the transverse
components of the magnetic field.

From the general formulas for the fields in Eq.~\eqref{Eq:EB} we see that in
general (i.e., for values $0 < \gamma < 1$) both the electric and the
magnetic field get such an additional contribution with opposite circular
polarization, such that none of the fields exhibits a circular polarization
of its transverse components. The relative strength between the
counter-circular electric and magnetic contribution and thus the relative
degree of ellipticity is determined by the value of $\gamma$. 

We also note that the terms giving rise to the ellipticity are of second order in the paraxial parameter $f$.  This is again in line with the fact that the transverse components should have only even order contributions in that parameter \cite{lax1975maxwell}. Summarizing our results for the longitudinal and transverse components, it turns out that in zeroth and first order of $f$ the fields are independent of the value of $\gamma$, i.e., on the gauge chosen to complement the transverse vector potential. All higher order terms will then depend on that choice. These differences will be therefore particularly important in the case of tightly focused beams, as will be discussed in more detail in Sect.~\ref{experiment}.

\subsubsection{Mixing of topological and spin indices}
\label{mixing}

Keeping in mind that $\sigma$ denotes the circular polarization and $\ell$
the topological charge of the transverse components of the original vector
potential, the transverse components of the electric and the magnetic fields
inherit a term with the same polarization and topological charge of $\mathbf
A_\perp$. However, in the {\it 2CA} case the electric field and in the {\it
3CA} case the magnetic field get additional terms with both circular
polarizations leading in general to an elliptic polarization. At the same
time, the azimuthal dependence of these fields is not anymore given by
$\exp(i\ell \varphi)$ but there are additional terms with a dependence
$\exp[i(\ell+\sigma \pm 1) \varphi]$. This shows that in the {\it 2CA} case
the transverse component of the electric field and in the {\it 3CA} case the
transverse components of the magnetic field are not anymore characterized by
a well-defined circular polarization and topological charge. Instead they
exhibit a mixture of orbital (index $\ell$) and spin (index $\sigma$) angular
momenta. Indeed, such a mixing of spin and orbital angular momentum has been found for tightly focused beams both in the case of Bessel \cite{bliokh2010angular} and LG \cite{iketaki2007inv, zhao2007spin, monteiro2009ang, bliokh2011spin} beams.

For $\sigma = 1$ there is one additional term with again $\sigma =
1$ and topological charge $\ell$ and another one with $\sigma = -1$ and
topological charge $\ell+2$. For $\sigma = -1$ there is again one additional
term with $\sigma = -1$ and topological charge $\ell$ and another one with
$\sigma = 1$ and topological charge $\ell-2$. This confirms that there is
indeed a coupling of SAM and OAM, since all terms have the same sum $\ell +
\sigma$ of topological charge and spin index.

\subsubsection{Spatial profiles and fields close to the phase singularity}

Due to the relation $J_{-\ell}(x) = (-1)^\ell J_{\ell}(x)$ that holds for
Bessel functions of integer order, the radial dependence of the fields is
determined by the modulus of the topological charge. This has led to a
classification of the beams into a {\it parallel class}, where
$\mathrm{Sign}(\ell)=\mathrm{Sign}(\sigma)$, and an {\it antiparallel class}
with $\mathrm{Sign}(\ell) \ne \mathrm{Sign}(\sigma)$
\cite{quinteiro2015formulation,quinteiro2017formulation}. In the parallel
class (and also in the case $\ell = 0$) the additional term with opposite
circular polarization has a modulus of the topological charge of $|\ell|+2$, i.e.,
the modulus is larger than in the terms with the original circular
polarization. Since close to the beam center Bessel functions behave like
$J_\ell(q_r r) \sim (q_r r)^{|\ell|}$, this means that the terms with
opposite circular polarization are usually negligible in this region. In
contrast, in the antiparallel class for beams with $|\ell| \ge 2$ the
additional term with opposite circular polarization has a modulus of the
topological charge of $|\ell|-2$, i.e., the modulus is smaller than in the
terms with the original circular polarization. Therefore, in the region close
to the beam center this term strongly dominates over the original one. In
particular, for $|\ell|=2$ the terms with opposite circular polarization have
a finite field strength at the beam center, in contrast to the terms with the
original circular polarization, which vanish according to $(q_r r)^{2}$. A
special case are antiparallel beams with $|\ell|=1$ where the terms with both
circular polarizations have the same radial dependence $\sim (q_r r)$.

In a previous work \cite{quinteiro2017formulation} we have studied the
behavior of Bessel beams from the antiparallel class close to the phase
singularity $r=0$. There we predicted the existence of an atypically strong
magnetic field. However, the origin of this asymmetry between the electric
and magnetic fields remained unclear. Based on the results of the present
paper, it is clear that this asymmetry was caused by the choice of the
potentials for the Bessel beam, which in the present notation were of {\it
3CA} type. Here we restore the symmetry by finding a procedure (i.e., using
{\it 2CA} potentials) that yields fields whose electric field exhibits the
same behavior close to the phase singularity and, moreover, by finding a
generalized procedure where this behavior can be found in both electric and
magnetic field.

\subsection{Laguerre-Gaussian beams}

Let us now apply the general formalism presented in Sect.~\ref{general} to the case of LG modes. The transverse components of the vector potential for a circularly polarized LG beam can be written as \cite{loudon2003theory}
\begin{equation}\label{Eq:A_LG}
\mathbf A_\perp(\mathbf r,t)
=  A_0   u_{q_z,\ell}(r,z) e^{i\ell \varphi}
e^{i (q_z z -\omega t)} {\boldsymbol \epsilon}_\sigma
\end{equation}
with $\omega=c q_z$ and the LG mode function $u_{q_z,\ell}(r,z)$ is given by
\begin{eqnarray}
  \hspace{-5mm} u_{q_z,\ell}(r,z)
&=&
  \frac{1}{\sqrt{\pi |\ell|!}} \left( \frac{\sqrt{2}}{w_0} \right)^{|\ell|+1}
  r^{|\ell|}
\nonumber \\
&&
  \times \exp{\left[ -\frac{r^2}{w_0^2}
  + \frac{2ikzr^2}{l^2}-2i(|\ell|+1)\frac{z}{l}\right]} .
\end{eqnarray}
Here, $w_0$ denotes the beam waist and $l=q_z w_0^2=2z_R$ is the diffraction length, $z_R$ being the Rayleigh range \cite{lax1975maxwell, loudon2003theory}. The paraxial parameter is given by $f=(q_z w_0)^{-1}$ leading to $(q_z l)^{-1}=f^2$. 

Again, the correction terms are related to the term $\nabla_\perp \cdot \mathbf{A}_\perp$. Using
Eq.~\eqref{Eq:A_LG}, this term reads
\begin{eqnarray}
  \nabla_\perp \cdot \mathbf A_\perp(\mathbf r, t)
&=&
  \frac{A_0}{\sqrt{2}}    \left( \frac{\partial u_{q_z,\ell}}{\partial r} -\sigma \ell \frac{u_{q_z,\ell}}{r} \right) \nonumber \\
  & & \times e^{i (\ell + \sigma) \varphi}
  e^{i (q_z z -\omega t)}.
\end{eqnarray}

As previously mentioned, the most common derivation leading to LG optical vortex fields is the one making use of the Lorenz gauge supplementing the transverse vector potential with a scalar potential; this is what here is called the \textit{2CA} procedure corresponding to $\gamma=0$ and $\overline{\gamma}=1$. According to Eq.~\eqref{Eq:2CA_Phi} the scalar potential is then given by  
\begin{eqnarray}
\Phi^{(0)}  (\mathbf r,t) &=& -i \frac{c^2}{\omega} \nabla_\perp \cdot \mathbf A_\perp(\mathbf r, t)\nonumber \\
&=&  -i \frac{c^2 A_0}{\sqrt{2}\omega}    \left( \frac{\partial u_{q_z,\ell}}{\partial r} -\sigma \ell \frac{u_{q_z,\ell}}{r} \right)  \nonumber \\
& & \times e^{i (\ell + \sigma) \varphi}
  e^{i (q_z z -\omega t)},
\end{eqnarray}
in agreement with the derivation in Ref.~\cite{loudon2003theory}. From this potential we obtain a longitudinal component of the electric field
\begin{equation}\label{Eq:Ez_Phi}
E_z^{(0)}(\mathbf r,t) = -\partial_z  \Phi^{(0)}  (\mathbf r,t) .
\end{equation}
This field has a first order contribution in the paraxial parameter $f$ given by $i q_z \Phi^{(0)}$ and corrections of third and higher orders in $f$, resulting from the $z$-dependence of the mode function $u_{q_z,\ell}$. The first order is again in agreement with Loudon's derivation \cite{loudon2003theory}. The higher order corrections are beyond the validity of the paraxial wave equation and have therefore been neglected in \cite{loudon2003theory}. The derivatives of $\Phi^{(0)}$ with respect to $x$ and $y$ lead to additional contributions in the transverse electric fields. Like in the case of Bessel beams they involve terms $\sim \exp[i(\ell+\sigma \pm 1)]$ and thus exhibit mixing of topological and spin indices. However, they are of second order in the paraxial parameter and therefore beyond the validity of the paraxial wave equation. Consequently they have been neglected in \cite{loudon2003theory}.

Let us now turn to the \textit{3CA} procedure corresponding to $\gamma=1$ and $\overline{\gamma}=0$. In this case  the scalar potential vanishes and the $z$-component of the vector potential is obtained from  
(see Eq.~\eqref{Eq:3CA_Az})
\begin{equation}
\partial_z A_z^{(1)} (\mathbf{r},t) =  -  \nabla_\perp \cdot \mathbf{A}_\perp(\mathbf{r},t).
\end{equation}
Here, the integration of this equation is more complicated than in the case of Bessel beams because the mode function  $u_{q_z,\ell}$ depends on $z$. However, this dependence occurs on a length scale of the diffraction length $l$, which is very slow compared to the variation of $\exp(iq_z z)$. Neglecting the $z$-dependence of  $u_{q_z,\ell}$, we obtain
\begin{eqnarray}
A_z^{(1)}  (\mathbf r,t) &=&  \frac{i}{q_z} \nabla_\perp \cdot \mathbf A_\perp(\mathbf r, t)\nonumber \\
&=&  -i \frac{A_0}{\sqrt{2} q_z}    \left( \frac{\partial u_{q_z,\ell}}{\partial r} -\sigma \ell \frac{u_{q_z,\ell}}{r} \right) \nonumber \\
& & \times e^{i (\ell + \sigma) \varphi}
  e^{i (q_z z -\omega t)}
\end{eqnarray}
and the corrections which have been neglected are of the order $f^2$. From this potential we now obtain a longitudinal component of the electric field
\begin{equation}\label{Eq:Ez_Az}
E_z^{(1)}(\mathbf r,t) = -\partial_t  A_z^{(1)}  (\mathbf r,t) .
\end{equation}
This field component is of first order in the paraxial parameter $f$ and furthermore it is in agreement with the corresponding field component obtained in the \textit{2CA} procedure. The next higher order corrections are of third order in $f$; these terms are different in the two procedures because in the \textit{2CA} case they result from the derivative of the mode function with respect to $z$ (see Eq.~\eqref{Eq:Ez_Phi}) while in the \textit{3CA} case they result from its integration with respect to $z$. These findings are again in agreement with the results of Lax \textit{et al.} \cite{lax1975maxwell} that the longitudinal components of the fields have only contributions of odd order in $f$ and the $f$-th order terms are completely determined by the $(f-1)$-th order terms of the transverse field components.

The derivatives of $A_z^{(1)}$ with respect to $x$ and $y$ lead to additional contributions to the transverse components of the magnetic field, as seen in Eq.~\eqref{Eq:EB}, exhibiting mixing of OAM and SAM. They are again of second order in $f$ and are therefore beyond the validity of the paraxial wave equation. 

We thus find that, much like in the case of Bessel beams, up to first order in the paraxial parameter the \textit{2CA} and \textit{3CA} procedures lead to the same electromagnetic fields. Beyond this level, i.e., in particular in tightly focused beams, differences will appear. However, while Bessel beams are correct solutions of Maxwelll's equations for arbitrary values of the paraxial parameter, LG beams are only correct up to first order. Therefore, to analyze differences in the fields of LG beams caused by the \textit{2CA} and \textit{3CA} procedure the description of these beams has to go beyond the paraxial regime, e.g., by including higher orders in the framework of the systematic approach discussed by Lax {\it{et al.}}~\cite{lax1975maxwell} or by studying tightly focused LG beams  \cite{iketaki2007inv, zhao2007spin, monteiro2009ang, bliokh2011spin} where indeed mixing of orbital and spin angular momentum has been found, as we have explicitly discussed for Bessel modes in Sect.~\ref{mixing}.

\section{Experimental considerations}\label{experiment}

We now discuss how one could experimentally distinguish between the different
fields considered above. We will concentrate on the case of Bessel beams because there the derived fields are exact solutions of the Maxwell equations. Measuring an optical field is usually done by
absorbing a photon in the detector. The strongest interaction is typically
achieved for electric dipole transitions, therefore here we concentrate on
measuring the electric field profiles of different beams. Let us start the
discussion by considering the required spatial scales.

As can be seen from Eqs.~\eqref{Eq:E_2CA_b} and \eqref{Eq:E_3CA_b}, the terms
which lead to the differences in the transverse components of $\mathbf{E}$
have a prefactor of $q_r^2/(q_r^2+q_z^2)$. With the beam waist $w=1/q_r$ and
the wavelength $\lambda=2\pi/\sqrt{q_r^2+q_z^2}$, the intensity of these
contributions is thus smaller by a factor of $(\lambda/2\pi w)^4$. Therefore,
to increase the sensitivity the beam should be focused as tightly as
possible. Let us assume that the beam is focused to $w=\lambda/2$, then the
characteristic scale for the variation of the field strength is a few hundred
nanometers. To measure these variations, a probe in the range of a few tens
of nanometers is required. This could be realized, e.g., by a semiconductor
nanostructure like a quantum dot \cite{quinteiro2014light} or a single
trapped ion \cite{Schmiegelow2016transfer}. In the following we will
concentrate on the latter scheme.

\begin{center}
\begin{figure}[t]
\centering
\includegraphics[width=0.98\columnwidth]{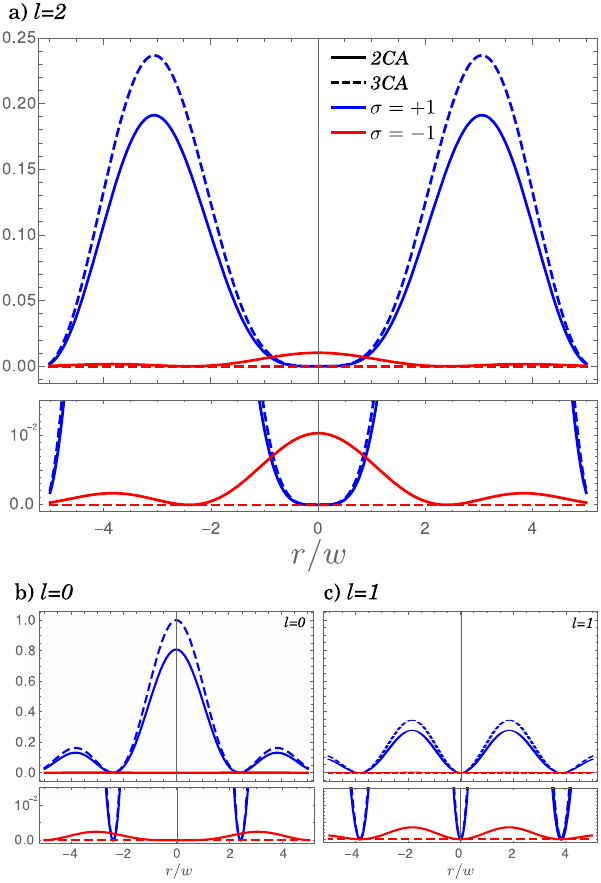}\\
\caption{Normalized transition probabilities as a function of the position of the ion for different beams all with waist $w=\lambda/2$. The probabilities are calculated for a dipole transition with change in magnetic quantum number $\Delta m =1$, assuming a negligible position spread for the center of mass of the ion. Panel a) refers to beams with chirality $\ell=2$, panels b) and c) refer to beams with  $\ell=1$ and $1$, respectively. Blue and red predictions are for the two different circular polarizations $\sigma =\pm 1$ while solid and dashed lines correspond to the prediction expected for a \textit{2CA} or \textit{3CA} model, respectively.
} \label{fig:profiles}
\end{figure}
\end{center}

We consider a single trapped ion interacting with a structured light beam as
in Ref.~\cite{Schmiegelow2016transfer}. The ion is laser cooled to an
average spread of 60~nm and can be positioned along the beam with
sub-nanometer precision. The beam shape can be measured by scanning the ion's
position across a beam while observing excitation probabilities on a given
atomic transition. Excitation probabilities, in this kind of experiments, are
usually measured by state dependent fluorescence~\cite{leibfried2003quantum}.

We focus here on the interaction with a dipole-allowed transition such as the S-P
transition in a one-electron atom. More specifically, we consider a Zeeman
split manifold such that one can independently observe absorption involving
different change in magnetic quantum number $\Delta m$. The strength of these
transitions will depend on the term accompanying each polarization in
Eq.~\eqref{Eq:E_2CA_b} for the electric
field \cite{quinteiro2017twisted, quinteiro2010ele}. For the case where the
external magnetic field and beam propagation direction are collinear, the
mapping is trivial: $\Delta m \longleftrightarrow \epsilon_{\Delta m} $. This
way, choosing the laser frequency and polarization one can probe the
different contributions of the electric field and discern whether the beam is of the \textit{2CA} or \textit{3CA} type.

In Fig.~\ref{fig:profiles} we show the expected results of a possible
experiment. The laser is tuned to a dipole-allowed transition with change in
total angular momentum $\Delta m=1$. The beam is tightly focused to
$\lambda/2$. The excitation probability cross sections is shown for the two
different polarizations and for three different spatial structures $l=0 , 1, 2$. The bottom rows present an expanded view of the low excitation range.

We suppose we can freely choose the beam's polarization and chirality but cannot assert whether it is of the type \textit{2CA} or \textit{3CA}. A detailed inspection of the expected excitation curves shown in Fig.~\ref{fig:profiles} shows that the best family of beams one can use to distinguish between \textit{2CA} and \textit{3CA} are those with $\ell=2$. For this family [Fig.~\ref{fig:profiles}a)] the expected spatial profile for $\sigma=+1$ (blue curves) has always a two peak structure with a pronounced zero at the center irrespective of the model. The reason is that here the spatial structure is always determined by the Bessel function $J_2$. However, when the polarization is changed to $\sigma=-1$ (red curves) the \textit{2CA} model predicts a three peak structure with a maximum at the center of the beam, which is caused by the contribution with the Bessel function $J_0$ arising from the mixing of topological and spin index, while the \textit{3CA} model predicts no interaction since here the polarization of the electric field remains purely $\sigma=-1$. In particular, the presence (or absence) of the peak at the center when changing the polarization from plus to minus would reveal if the beam is of the \textit{3CA} (or \textit{2CA}) type.

It is important to note that polarizations can never be set perfectly, so measuring small contributions when there is another stronger competing one is always challenging. For example, in the case just described, a small amount of the wrong polarization would produce a spurious signal with a spatial pattern corresponding to that polarization. For this reason it is important to choose profiles which have distinct features for each polarization. The case of $\ell=1$ [Fig.~\ref{fig:profiles}c)] is an example of how no shape difference for the interactions with different polarizations makes this family of beams a poor candidate for the method described above. The reason is that here the mixing of topological and spin index couples Bessel functions $J_1$ and $J_{-1}$ which, however, have the same spatial profile.

The $\ell=0$ case [Fig.~\ref{fig:profiles}b)], in turn, provides some distinguishability but not as good as the $\ell=2$ case first discussed. Here, the $\sigma=-1$ beam in the \textit{2CA} case acquires a contribution with the spatial profile of the Bessel function $J_2$, which has non-vanishing values at positions where the $\sigma=+1$ beam is zero. Again, the excitation probability of the $\sigma=-1$ beam in the \textit{3CA} case is strictly zero because of the absence of mixing between spin and orbital angular momentum.

\section{Conclusions}\label{conclusions}

We have analyzed different methods to derive electromagnetic fields from an ansatz for the  transverse vector potential. These methods were inspired by the ones normally used to derive LG and Bessel beams.  In order to compare the methods we applied both methods to both types of field modes, LG and the Bessel modes. We discussed in detail the case of Bessel beams, which was motivated by the fact that these are solutions of the full Maxwell equations and not limited to the paraxial approximation. We showed that the procedures lead to different electromagnetic fields for the same ansatz. These two types of beams, which we call circular electric and circular magnetic, exhibit different properties that stem from the presence of an anomalous term in the transverse part of either the electric or magnetic fields. We then showed that the two procedures can be regarded as limiting cases of a generalized procedure, which interpolates smoothly between the limiting cases and results in a class of electromagnetic fields, that have the anomalous term in both electric and magnetic fields. The same behavior has been found for LG beams; however, since the anomalous terms are of second order in the paraxial parameter LG beams beyond the paraxial limit have to be considered, either by including higher order terms in the paraxial expansion or by studying tightly focused beams.

The second order terms in the paraxial parameter in the transverse components of the electric and magnetic field of both Bessel and LG beams are responsible for the mixing of orbital and spin angular momentum. This becomes more important when beams are strongly focused as has already been pointed out in the literature \cite{monteiro2009ang, bliokh2010angular, bliokh2011spin, iketaki2007inv}. Here we provided an explicit form for this mixing term and we demonstrated that it can be present only in the electric field, only in the magnetic field, or with an arbitrary weight in both fields.

\section{Acknowledgment}\label{ack}

G.\ F.\ Q.\ acknowledges support from Agencia Nacional de Promoci\'on Cient\'ifica y Tecnol\'ogica (grant PICT 2013-0592), from the german Deutsche Akademische Austauschdienst (DAAD) and argentine CONICET. C.\ T.\ S.\ acknowledges support from grant PICT 2014-3711.


\bibliography{std,twistedlight_v2}

\end{document}